\documentclass[11pt,twoside,usenatbib,useAMS]{article}

\usepackage{asp2006}
\usepackage{epsf}
\usepackage{epsfig}
\usepackage{lscape}

\def \hii{H{\scriptsize II}}

\def \nhthree{\mbox{NH$_3$}}
\def \nhone{\mbox{NH$_3$(1,1)}}

\def \nhfour{\mbox{NH$_3$(4,4)}}

\begin{document}

\title{Determining the relative evolutionary stages of very young massive star formation regions} 
\author{S.N.Longmore\altaffilmark{1,2,3,7}, M.G.Burton\altaffilmark{1}, C.R.Purcell\altaffilmark{4}, P.Barnes\altaffilmark{5} \& J.Ott\altaffilmark{6} } 
\altaffiltext{1}{School of Physics, University of New South Wales, Kensington, NSW 2052, Sydney, Australia} 
\altaffiltext{2}{Australia Telescope National Facility, CSIRO, PO Box 76, Epping, NSW 1710, Australia}
\altaffiltext{3}{Harvard-Smithsonian Center for Astrophysics, 60
Garden Street, Cambridge, MA 02138, USA} 
\altaffiltext{4}{University of Manchester, Jodrell Bank Observatory, Macclesfield, Cheshire SK11 9DL, UK} 
\altaffiltext{5}{School of Physics A28, University of Sydney, NSW 2006, Australia} 
\altaffiltext{6}{National Radio Astronomy Observatory, 520 Edgemont Road, Charlottesville, VA 22903, USA}
\altaffiltext{7}{E-mail: slongmore@cfa.harvard.edu}

\begin{abstract} 

We have recently completed an observing program with the Australia
Telescope Compact Array towards massive star formation regions traced
by 6.7\,GHz methanol maser emission. We found the molecular cores
could be separated into groups based on their association with/without
methanol maser and 24\,GHz continuum emission. Analysis of the
molecular and ionised gas properties suggested the cores within the
groups may be at different evolutionary stages. In this contribution
we derive the column densities and temperatures of the cores from the
$\nhthree$ emission and investigate if this can be used as an
indicator of the relative evolutionary stages of cores in the sample.

The majority of cores are well fit using single-temperature large
velocity gradient models, and exhibit a range of temperatures from
$\sim$10\,K to $>$200\,K. Under the simple but reasonable assumption
that molecular gas in the cores will heat up and become less quiescent
with age due to feedback from the powering source(s), the molecular
gas kinetic temperature combined with information of the core
kinematics seems a promising probe of relative core age in the
earliest evolutionary stages of massive star formation.

\end{abstract}



\section{Introduction}

Although short in astronomical terms, the formation time-scale for
massive stars is long enough that observations of an individual
massive star formation (MSF) region can only provide a snap-shot of
the formation process. Given the paucity of young Galactic MSF
regions, developing an end-to-end model of the formation process will
be greatly facilitated by the ability to accurately determine the
relative evolutionary stage of different regions. Observationally, MSF
regions tend to be placed into broadly defined groups such as
ultra-compact and hyper-compact \hii~regions and hot/cold molecular
cores. However, it is not clear if the discrete nature of the
formation process implied by this grouping is a physical reality
(suggesting each of currently defined groups represent relatively
long-lived stages which quickly transition to the next stage), or if
the formation process is in fact more continuous.

In an attempt to develop our understanding of the early evolutionary
stages of the MSF process, \citet{L07A} (L07A) have completed an
observing program with the Australia Telescope Compact Array (ATCA)
towards MSF regions traced by 6.7\,GHz methanol maser emission (a
powerful sign-post of very early massive star formation). The aim of
these observations was to derive the physical properties of the gas in
these regions and see how accurately it is possible to determine the
relative evolutionary stages of the cores. The initial results
suggested cores could be separated into different evolutionary stages
based on their association with/without methanol maser and 24\,GHz
continuum emission. In this contribution we focus on deriving the core
temperatures and investigate the use of molecular gas kinetic
temperature as an indicator of the cores' evolutionary stage.

\section{Deriving properties of molecular gas from $\nhthree$ observations}
\label{sec:deriving_gas_properties}

\subsection{Observations}
\label{sub:observations}
Observations of $\nhone$, (2,2), (4,4) \& (5,5) and 24\,GHz continuum
were carried out using the ATCA towards 21 MSF regions traced by
6.7\,GHz methanol maser emission. The H168 [$\nhone$ \& (2,2) observed
simultaneously] and H214 [$\nhfour$ \& (5,5) observed simultaneously]
antenna configurations with both East-West and North-South baselines,
were used to allow for snapshot imaging. Primary and characteristic
synthesised beam sizes were $\sim$2.$\arcmin$2 and $\sim$$8 - 11
\arcsec$ respectively. Characteristic spectra were extracted at every
transition for each core at the position of the peak $\nhone$ emission
and fit using CLASS. The $\nhthree$ column density and gas kinetic
temperature were calculated from the L07A data using the standard
analysis procedure \citep[see e.g.][]{ho_townes1983,
ungerechts1986}. In $\S$~\ref{sub:standard_analysis}, we assess the
validity of the assumptions used in this procedure for the L07A
sample. $\S$~\ref{sub:lvg_modelling} outlines the large velocity
gradient models used to derive gas kinetic temperatures.

\subsection{Assessing the assumptions used in the standard $\nhthree$ analysis}
\label{sub:standard_analysis}
The three main underlying assumptions in the standard analysis used to
derive $\nhthree$ column density and molecular gas kinetic temperature
are that:

\begin{enumerate}
  \item the excitation temperature is the same for all hyperfine
  components within an inversion transition.
  \item the excitation temperature is the same for all inversion
  transitions.
  \item the emission from all inversion transitions comes from the
  same gas so has equal line width and beam filling factor.
\end{enumerate}

\noindent
The validity of these assumptions for the L07A data are assessed below.

\paragraph{Assumption 1}
Given the extreme $\nhone$ hyperfine anomalies in the outer satellite
lines of some L07A spectra, the first assumption is clearly invalid
for these outer satellites. The optical depth derived from CLASS
(which fits to all 18 components under the equal excitation
temperature assumption) is therefore potentially incorrect. As
hyperfine structure is not detected in the other inversion
transitions, it is not possible to check for further anomalies in the
excitation conditions.

\paragraph{Assumption 2}
Without detected hyperfine structure in the energy levels above
$\nhone$, it is not possible to measure the excitation temperature
within other inversion transitions to test the second
assumption. Given the very similar morphology of the $\nhone$ and
(2,2) emission, this lends support to the idea that the excitation
conditions are similar for these two transitions. However, the same
cannot be said for the $\nhfour$ \& (5,5) emission [as this is always
unresolved], rendering the second assumption suspect for these cores.

\paragraph{Assumption 3}
As the $\nhone$ and (2,2) emission is generally somewhat extended with
very similar morphology, the assumption that emission in these
transitions is from the same gas seems valid. However, as the
$\nhfour$ \& (5,5) emission is always unresolved, it is not possible
to test this assumption for these transitions.

\vspace{2mm}
\noindent
In summary, the standard analysis:

\begin{itemize}
  \item will be reliable for sources detected only in $\nhone$ and
  (2,2) with no $\nhone$ hyperfine anomalies
  \item will introduce an uncertainty in the derived optical depth for
  sources detected in only $\nhone$ and (2,2) with $\nhone$ hyperfine
  anomalies
  \item may not be reliable for sources with detected $\nhfour$ and
  (5,5) emission. Modelling is required to derive accurate properties
  of the gas for these sources.
\end{itemize}

\subsection{LVG modelling}
\label{sub:lvg_modelling}
Having derived a column density for each detected transitions for all
the cores, we used the large velocity gradient (LVG) models described
in Ott et al. 2005 to derive the core kinetic temperatures. Assuming a
velocity gradient of 1 kms$^{-1}$pc$^{-1}$, background temperature of
2.73\,K and hydrogen volume density of 10$^5$cm$^{-3}$, $\chi^2$
minimisation was used to calculate the best-fit kinetic temperature
and total ammonia column density to the measured column density within
each transition.

\section{Results}
\label{sec:results}

\begin{figure*} 
 \begin{center} 
  \includegraphics[width=6.5cm, angle=-90, trim=0 0 -5 0]{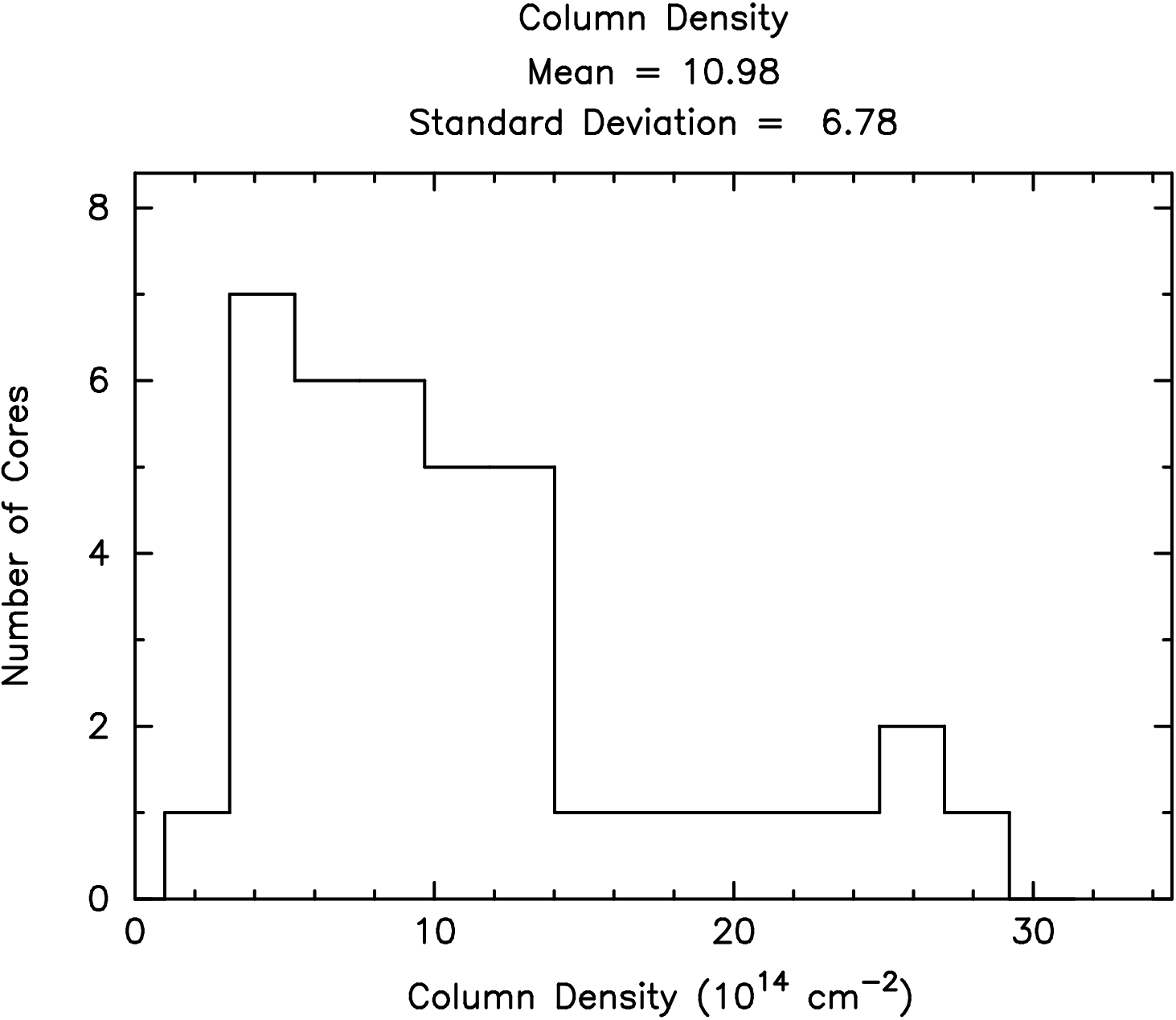} 
\end{center}
  \caption[Column density histogram]{{\small Histogram of the total
  $\nhthree$ column density. }}
           
  \label{fig:col_dens_hist}
\end{figure*}

\paragraph{Column Density}
Figure~\ref{fig:col_dens_hist} shows the histogram of the total
$\nhthree$ column density derived using the standard $\nhthree$
analysis procedure. The mean value of $\sim$10$^{15}$\,cm$^{-2}$ is
similar to values derived for other high-mass cores.

\begin{figure*} 
\begin{center} 
\begin{tabular}{cc} 
 \includegraphics[width=7.5cm, height=7.5cm, angle=-90, trim=0 0 -5 0]{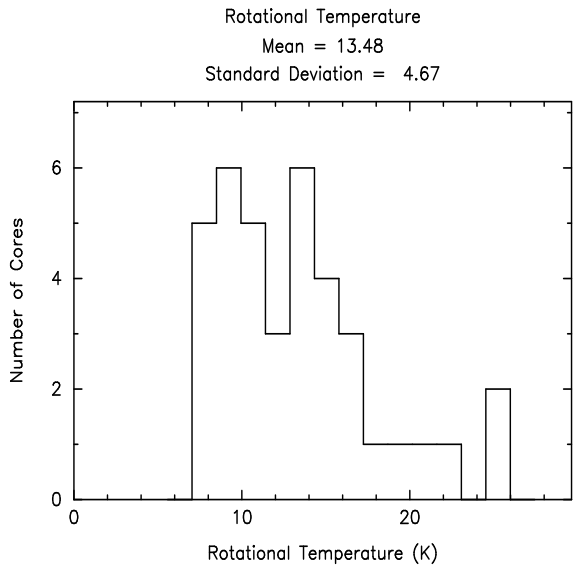} & 
\end{tabular}
\end{center}
\caption{Histograms of the molecular gas temperature derived from the
cores with $\nhone$ \& (2,2) emission. }
\label{fig:trot_hist}
\end{figure*}

\paragraph{Temperature}
Figure~\ref{fig:trot_hist} shows a temperature histogram for cores
with only detected $\nhone$ \& (2,2) emission (for which the
assumptions in the standard analysis should be reliable). Given that
this includes cores with large $\nhone$ hyperfine anomalies
(potentially introducing uncertainties in the $\nhone$ column density
and hence temperature), the spread in temperature for these cores is
relatively small.

\begin{figure*} 
 \begin{center} 
  \includegraphics[width=6.5cm, angle=-90, trim=0 0 -5 0]{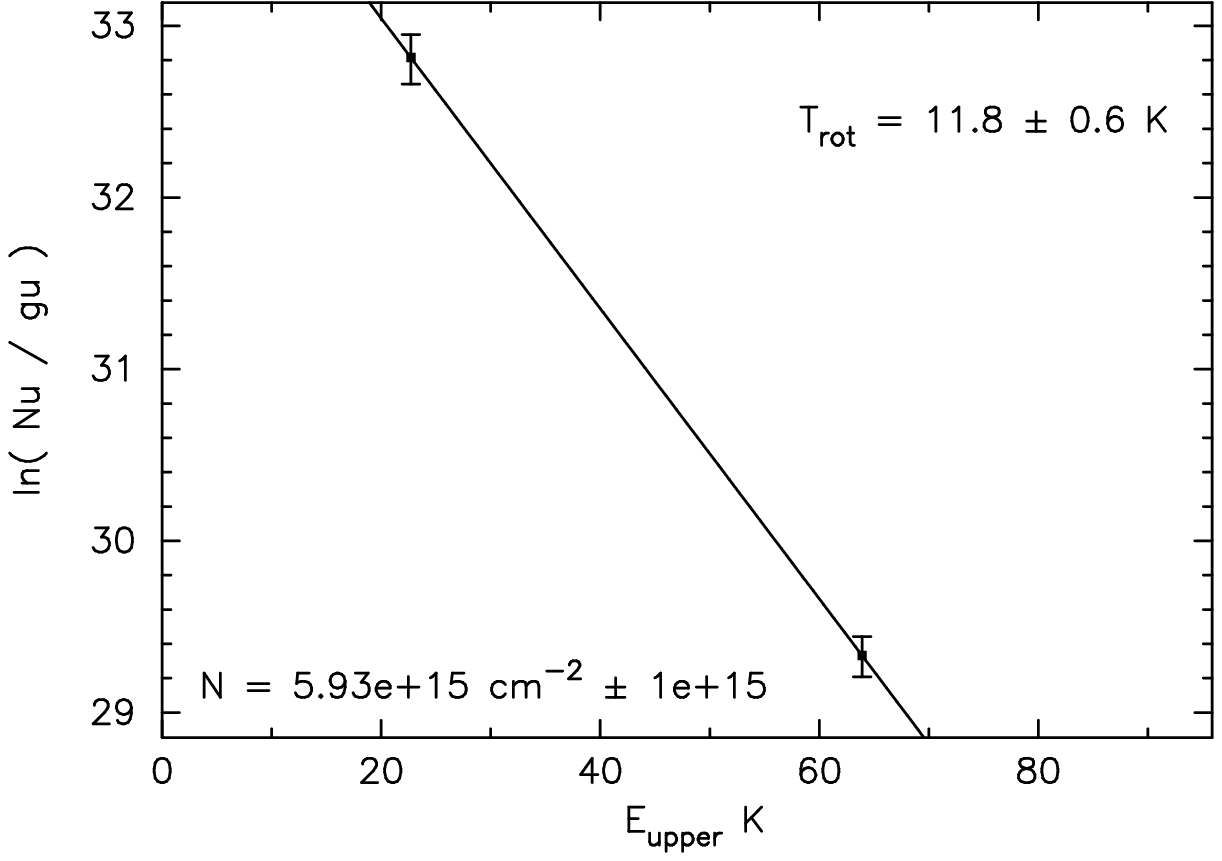} 
  \includegraphics[width=6.5cm, angle=-90, trim=0 0 -5 0]{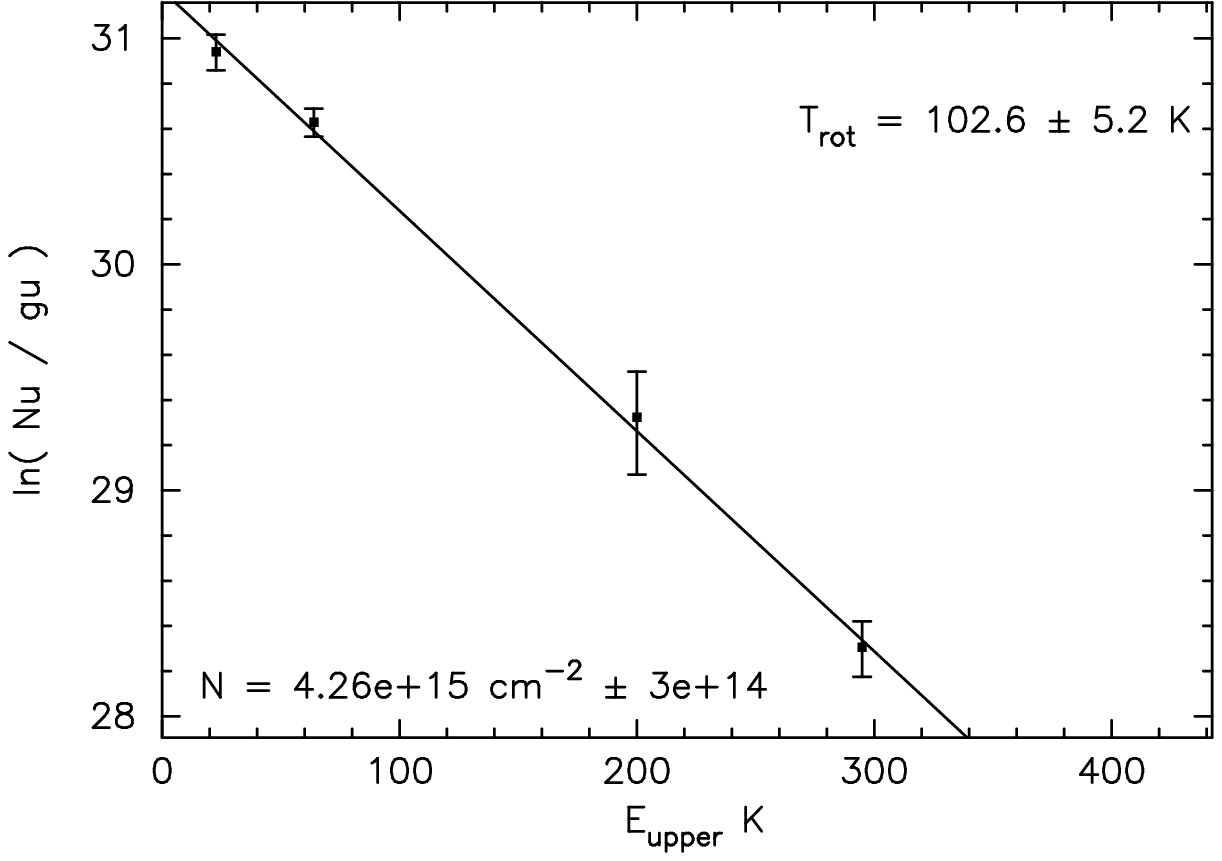}  
  \includegraphics[width=6.5cm, angle=-90, trim=0 0 -5 0]{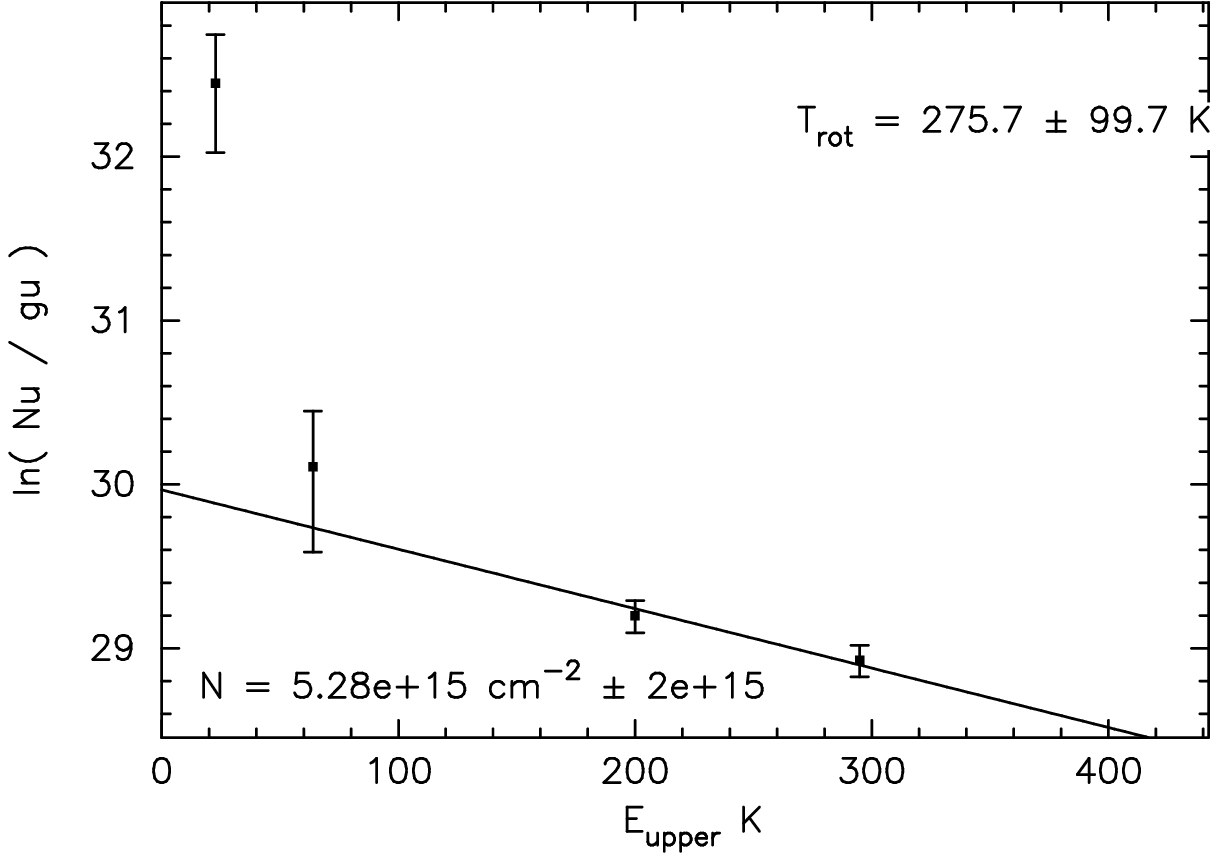}  
\end{center}
  \caption[Example $\nhthree$ rotational diagrams]{{\small Example
           $\nhthree$ rotational diagrams for three cores, plotting
           the column density derived for each transition as a
           function of the transition energy. The straight line shows
           the weighted best-fit line through the points. The
           temperature is proportional to the inverse of the line
           gradient (the value of which is shown in the top-right of
           each plot) and the intercept point on the y-axis gives the
           total $\nhthree$ column density (the value of which is
           shown in the bottom-left of each plot). The core in the top
           plot is only detected at $\nhone$ \& (2,2) while the cores
           in the middle and bottom plot are detected from $\nhone
           \rightarrow (5,5)$. The cores in the top and middle plots
           are well fit by a single temperature model but the single
           temperature approximation is a poor fit to the core in the
           bottom plot. As discussed in $\S$\ref{sec:results}, the
           emission is likely to be coming from gas at two different
           temperatures. Further modelling is required to determine
           the core structure.}}
           
  \label{fig:rot_diags}
\end{figure*}

The temperature derivation for cores with detected $\nhfour$ \& (5,5)
emission is more complicated. The single-temperature LVG model fits
well to most cores but poorly to others. To illustrate this,
Figure~\ref{fig:rot_diags} shows rotational diagrams (effectively
column density within each transition vs. transition energy) for three
example cores, with the best-fit temperature plotted as a straight
line. Both of the cores in the top [only detected at $\nhone$ \&
(2,2)] and middle plots are well fit with a single temperature
component. Conversely, the column densities calculated for the core in
the bottom plot are clearly poorly fit with a single temperature
component. However, as described in $\S$~\ref{sub:observations}, the
$\nhone$ \& (2,2) transitions were observed using a different array
configuration and set-up to the $\nhfour$ \& (5,5) transitions. It is
therefore possible that the bad fit to a single temperature component
is an artefact of comparing gas properties derived from the different
observing set-ups. Firstly, comparing column densities between these
two line pairs may suffer from systematic offsets caused by the
$\sim$10\% uncertainty in the absolute flux calibration that doesn't
affect the simultaneously observed line-pairs. Similarly, the slightly
different array configurations used to observe the two line-pairs may
be introducing an offset in the derived column densities as each
observation samples slightly different spatial structure. However, as
each of the line-pairs were observed simultaneously, they sample the
same spatial structure and have the same absolute flux calibration
error. As the temperature between two transitions is calculated from
the ratio of measured intensity (rather than the absolute intensity
used to derive the column density) temperatures derived between
$\nhone$ \& (2,2) [T$_{12}$] and $\nhfour$ \& (5,5) [T$_{45}$] will
not suffer these problems. Additionally, the line-pairs have similar
excitation temperatures (23 \& 64\,K vs 200 \& 295\,K for $\nhone$ \&
(2,2) vs $\nhfour$ \& (5,5), respectively) so are much more likely to
be sampling the same gas. Deriving temperatures for these line-pairs
suggests there are two seperate components -- a cold component traced
by T$_{12}$ and a hot component traced by T$_{45}$. Comparing the
observed morphology of the line-pairs (the $\nhfour$ \& (5,5) emission
always being unresolved at the central peak of the generally extended
$\nhone$ \& (2,2) emission) further points to the emission from these
line-pairs coming from different gas. If so, this renders assumptions
2 and 3 in $\S$\ref{sub:standard_analysis} (used to derive the
$\nhfour$ \& (5,5) column densities) invalid. In conclusion, while it
seems likely there are multiple temperature components in these cores,
further analysis (e.g. radiative transfer modelling to fit the
physical structure of the core to the observed spectra) is required to
determine how this relates to the core structure. Therefore, in the
remaining sections we limit ourselves to discussing the general spread
in the core temperatures rather than focusing on individual
temperature measurements themselves.

\section{Discussion}
\label{sec:discussion}

\begin{figure}
    \begin{center}
      	  \includegraphics[width=5cm, angle=-90, trim=0 0 -5 0]{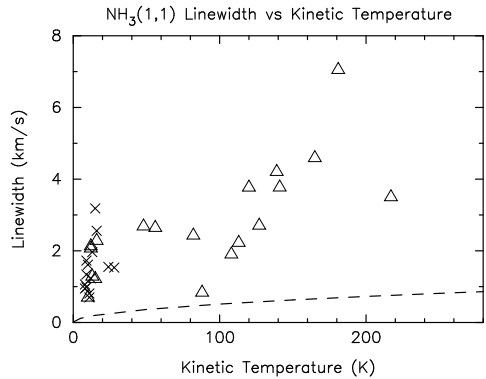}
    \end{center}
    \caption{$\nhone$ linewidth vs gas kinetic temperature. Cores with
    $\nhthree$ emission only (Group 1 in L07A) are shown as crosses
    while those with $\nhthree$ and methanol maser emission (Groups 2
    and 3 in L07A) are shown as triangles. The dashed line shows the
    expected linewidth due to purely thermal motions.}
	\label{fig:lw_temp}
\end{figure}

Analysis of the molecular/ionised gas in L07A suggested cores grouped
according to their association with/without 6.7\,GHz methanol maser
and 24\,GHz continuum emission were potentially at different
evolutionary stages. Making the reasonable assumption that cores heat
up and becomes less quiescent with age, we now investigate how the
derived molecular gas temperature fits in with the proposed
evolutionary scenario. As the core sizes are similar, L07A used the
$\nhone$ linewidths as a reliable indicator of how quiescent the gas
is, without worrying about its dependence on the core size
\citep{larson1981}. Figure~\ref{fig:lw_temp} [from \citet{L07B}] shows
$\nhone$ core linewidth vs kinetic temperature. Triangles and crosses
show sources with/without methanol maser emission, respectively. The
molecular gas in sources with methanol maser emission (triangles) are
generally significantly warmer than those without (crosses). However,
there are also a small number of cores with methanol maser emission
that have very cool temperatures and quiescent gas, similar to the
$\nhthree$-only cores. Modelling shows pumping of 6.7\,GHz methanol
masers requires local temperatures sufficient to evaporate methanol
from the dust grains (T$>\sim$90K) and a highly luminous source of IR
photons [\cite{cragg2005}] i.e.  an internal powering source. It is
therefore plausible that the cold, quiescent sources with methanol
maser emission are cores in which the feedback from the powering
sources have not had time to significantly alter the larger scale
properties of the gas in the cores. It is worth noting that
\citet{krumholz2006} predicts accretion luminosity at the earliest
stages of core collapse can heat the inner regions to 100\,K. This
would raise the characteristic fragmentation mass above that in the
outer regions, effectively suppressing fragmentation in the inner
regions. The sources we observe with potentially more than one
temperature component are therefore interesting targets for testing
the competitive accretion vs. turbulent core models of the formation
process.

Finally, concentrating on cores with a well defined temperature, we
note that: 1) there are a large number of cold (10 to 20\,K) cores
with quiescent gas which may be regions at the very earliest
evolutionary stages before any star-formation feedback has effected
the natal molecular material, and 2) the spread in core temperature
from $\sim$20 to 200\,K is continuous, suggesting the transition from
`cold' to `hot' core is not discrete. This urges a note of caution
when comparing samples of hot and cold cores or assuming a single dust
temperature when deriving properties for numerous sources from mm
continuum emission.

\section{Conclusions}
\label{sec:conclusions}
Based on the $\nhthree$ observations of \citet{L07A}, we have
performed a preliminary analysis of core temperature towards a sample
of very young MSF regions, to try and determine their relative
evolutionary stages. We find,
\begin{itemize}
  \item a large number of cold ($\sim$ 10\,K) cores with very
  quiescent gas. These are likely to be MSF regions at the earliest
  evolutionary stages.
  \item above $\sim$20\,K the cores exhibit a smooth range in
  temperature suggesting the transition from `cold' to `hot' cores is
  continuous rather than discrete.
  \item two lines of evidence suggesting some of the cores have more
  than one temperature component. Although more detailed modelling is
  required to accurately derive the core structure, these are
  potentially interesting for future study as regions in which the
  central powering sources has only recently begun to heat up its
  surrounding environment.
\end{itemize}

\noindent
In summary, molecular gas kinetic temperature combined with
information of the core kinematics appears to be a promising way to
compare the evolutionary state of very young MSF regions.

\acknowledgements

We would like to thank Thushara Pillai for her instructive
comments. SNL acknowledges support through a School of Physics
scholarship at UNSW. The Australia Telescope is funded by the
Commonwealth of Australia for operation as a National Facility managed
by CSIRO. This research has made use of NASA's Astrophysics Data
System. We also thank the Australian Research Council for funding
support from grant number DP0451893.

\bibliography{longmore_heidelberg}

\begin{thebibliography}{}

\bibitem[\protect\citeauthoryear{{Cragg}, {Sobolev} \& {Godfrey}}{{Cragg}
  et~al.}{2005}]{cragg2005}
{Cragg} D.~M.,  {Sobolev} A.~M.,    {Godfrey} P.~D.,  2005, \mnras, 360, 533

\bibitem[\protect\citeauthoryear{{Ho} \& {Townes}}{{Ho} \&
  {Townes}}{1983}]{ho_townes1983}
{Ho} P.~T.~P.,  {Townes} C.~H.,  1983, \araa, 21, 239

\bibitem[\protect\citeauthoryear{{Krumholz}}{{Krumholz}}{2006}]{krumholz2006}
{Krumholz} M.~R.,  2006, \apjl, 641, L45

\bibitem[\protect\citeauthoryear{{Larson}}{{Larson}}{1981}]{larson1981}
{Larson} R.~B.,  1981, \mnras, 194, 809

\bibitem[\protect\citeauthoryear{{Longmore}, {Burton}, {Barnes}, {Wong},
  {Purcell} \& {Ott}}{{Longmore} et~al.}{2007a}]{L07A}
{Longmore} S.~N.,  {Burton} M.~G.,  {Barnes} P.~J.,  {Wong} T.,  {Purcell}
  C.~R.,    {Ott} J.,  2007a, \mnras, 379, 535

\bibitem[\protect\citeauthoryear{{Longmore}, {Burton}, {Barnes}, {Wong},
  {Purcell} \& {Ott}}{{Longmore} et~al.}{2007b}]{L07B}
{Longmore} S.~N.,  {Burton} M.~G.,  {Barnes} P.~J.,  {Wong} T.,  {Purcell}
  C.~R.,    {Ott} J.,  2007b, in {Baan} W.,  {Chapman} J.,  eds, IAU Symposium
  242, Astrophysical masers and their environments, {The molecular environment
  of massive star forming cores associated with Class II methanol maser
  emission}

\bibitem[\protect\citeauthoryear{{Ungerechts}, {Winnewisser} \&
  {Walmsley}}{{Ungerechts} et~al.}{1986}]{ungerechts1986}
{Ungerechts} H.,  {Winnewisser} G.,    {Walmsley} C.~M.,  1986, \aap, 157, 207

\end{thebibliography}

\end{document}